\documentclass[fleqn,usenatbib]{mnras}

\usepackage{newtxtext,newtxmath}

\usepackage[T1]{fontenc}

\DeclareRobustCommand{\VAN}[3]{#2}
\let\VANthebibliography\thebibliography
\def\thebibliography{\DeclareRobustCommand{\VAN}[3]{##3}\VANthebibliography}

\usepackage{graphicx}	
\usepackage{amsmath}	
\usepackage{multirow}
\usepackage{ulem}

\usepackage{lipsum}

\newcommand{\kms}{\ifmmode{~{\rm km\,s}^{-1}}\else{~km~s$^{-1}$}\fi}

\title[The post-RGB post-CE central stars of the PN~Ou~5]{The post-common-envelope binary central star of the planetary nebula Ou~5: a doubly-eclipsing post-red-giant-branch system}

\author[D.\ Jones et al.]
{David Jones,$^{1,2}$\thanks{Email: djones@iac.es}
James Munday,$^{3}$
Romano L.M.\ Corradi,$^{4,1}$
Pablo Rodr\'iguez-Gil,$^{1,2}$
\newauthor
Henri M.J.\ Boffin,$^{5}$
Jiri Zak,$^{5}$
Paulina Sowicka,$^6$
Steven G.\ Parsons,$^7$
Vik S.\ Dhillon,$^{7,1}$
S.P.\ Littlefair$^7$
\newauthor
T.R.\ Marsh,$^3$
Nicole Reindl,$^8$
and Jorge Garc\'ia-Rojas$^{1,2}$
\\
$^{1}$Instituto de Astrof\'isica de Canarias, E-38205 La Laguna, Spain 
\\
$^{2}$Departamento de Astrof\'isica, Universidad de La Laguna, E-38206 La Laguna, Spain\\
$^{3}$Department of Physics, Gibbet Hill Road, University of Warwick, Coventry CV4 7AL, United Kingdom\\
$^{4}$GRANTECAN, Cuesta de San Jos\'e s/n, E-38712, Bre\~na Baja, La Palma, Spain\\
$^{5}$European Southern Observatory, Karl-Schwarzschild-str. 2, D-85748 Garching, Germany\\
$^{6}$Nicolaus Copernicus Astronomical Center, Bartycka 18, PL-00-716 Warsaw, Poland\\
$^{7}$Department of Physics and Astronomy, University of Sheffield, Sheffield, S3 7RH, UK\\
$^{8}$Institute for Physics and Astronomy, University of Potsdam, Karl-Liebknecht-Str. 24/25, D-14476 Potsdam, Germany
}

\date{Accepted XXX. Received YYY; in original form 2021 November 22}

\pubyear{2021}

\begin{document}
\label{firstpage}
\pagerange{\pageref{firstpage}--\pageref{lastpage}}
\maketitle

\begin{abstract}
We present a detailed study of the stellar and orbital parameters of the post-common envelope binary central star of the planetary nebula Ou~5.  Low-resolution spectra obtained during the primary eclipse -- to our knowledge the first isolated spectra of the companion to a post-common-envelope planetary nebula central star -- were compared to catalogue spectra, indicating that the companion star is a late K- or early M-type dwarf. Simultaneous modelling of multi-band photometry and time-resolved radial velocity measurements was then used to independently determine the parameters of both stars as well as the orbital period and inclination.  The modelling indicates that the companion star is low mass ($\sim$0.25~M$_\odot$) and has a radius significantly larger than would be expected for its mass.  Furthermore, the effective temperature and surface gravity of nebular progenitor, as derived by the modelling, do not lie on single-star post-AGB evolutionary tracks, instead being more consistent with a post-RGB evolution. However, an accurate determination of the component masses is challenging.  This is principally due to the uncertainty on the locus of the spectral lines generated by the irradiation of the companion's atmosphere by the hot primary (used to derive companion star's radial velocities), as well as the lack of radial velocities of the primary.
\end{abstract}




\section{Introduction}

The physical parameters of post-common-envelope (post-CE) binary stars offer some of the most stringent tests of our understanding of the CE \citep[e.g.,][]{toonen13,iaconi19,politano21}.  Similarly, post-CE central stars of planetary nebulae (PNe) are ideal for this purpose as the presence of the surrounding short-lived nebula ensures that the system is fresh out of the CE and has not yet had time to evolve/relax appreciably \citep{jones17,boffin19}.  Unfortunately, the nebula can also significantly complicate the situation, contaminating the photometry \citep[e.g.,][]{jones14,jones15} and/or preventing the accurate measurement of radial velocities \citep{miszalski11}.  Additionally, most post-CE central stars with a main-sequence companion \citep[as opposed to a second degenerate star;][]{boffin12} also exhibit extremely large irradiation effects which can be challenging to model using classical bolometric prescriptions \citep[][and references therein]{barman04,horvat19}.

The central star of the PN Ou~5 ($\alpha=21^h14^m20.03^s$ $\delta=+43^\circ41' 36.00''$, PN~G086.9$-$03.4) was shown by \citet{Ou52014Corradi} to be an eclipsing post-CE binary with one of the largest observed irradiation effects.  Those authors noted that this makes Ou~5 an interesting candidate for follow-up modelling as the eclipsing nature of the binary means that the stellar radii can be unambiguously derived and, while the deep primary eclipse shows that the central star is significantly brighter than the companion, the large irradiation effect means that the companion's radial velocities can be measured (potentially making Ou~5 a double-lined spectroscopic binary, providing the nebular contamination of the primary's absorption spectra is not too significant).  The nebula itself has a remarkable morphology, seemingly comprising barrel-shaped nested lobes \citep{Ou52014Corradi}, which is strikingly similar to the hydrodynamic simulations of post-CE PNe of \citet[model A2]{garcia-segura18}.

\citet{corradi15} demonstrated that the nebula of Ou~5 presents one of the largest known abundance discrepancy factors (with recombination line abundances of O$^{2+}$ exceeding collisionally-excited line abundances of the same ion by a factor of more than fifty), associated with the presence of an additional low-temperature, high-metallicity gas phase in the nebula, the origins of which are almost certainly related to the CE evolution of the central star \citep{jones16,wesson18}.  \citet{corradi15} also note the generally unusual nebular abundances -- high He/H, low N/O and low N/H -- which are not really consistent with either being type \textsc{i} or type \textsc{ii} \citep{peimbert78,faundez-abans87}.  However, it is important to keep in mind that the abundances in such high abundance discrepancy nebulae are difficult to constrain \citep[as it is near impossible to constrain the fraction of H in each of the gas phases;][]{gomez-llanos20}.

In this paper, we present extensive follow-up photometry and spectroscopy of the central stars of Ou~5 with the aim of characterising the orbital and stellar parameters.  In Sec.\ \ref{sec:obs}, observations, data reduction and initial analyses are outlined, while in Sec.\ \ref{sec:phoebe} the simultaneous modelling of the light and radial velocity curves is described, before concluding in Sec.\ \ref{sec:conc}.

\section{Observations}
\label{sec:obs}
\subsection{Photometry}
\label{sec:phot}

The $i$-band photometry presented in \citet{Ou52014Corradi}\footnote{Namely IAC80-CAMELOT, INT-WFC and WHT-ACAM images available in their respective archives, which were re-reduced following the routines described here.} was supplemented by additional observations, with the intention of either characterising the variability in other bands or better constraining the depth of the primary eclipse.

Time-series $i$-band photometry of the central star of Ou~5 was obtained with the Auxiliary-port CAMera (ACAM) mounted on the 4.2m William Herschel Telescope (WHT) on 18 August 2014 and 27 July 2015 with integration times of 60s. Further $g$- and $r$-band images were obtained on 3 September 2016 with integration times ranging from 45--300s depending on the filter and the weather conditions\footnote{For individual exposure times, see the ING archive at \url{https://casu.ast.cam.ac.uk/casuadc/ingarch/}}.

\begin{figure*}
    \centering
    \includegraphics[width=\textwidth]{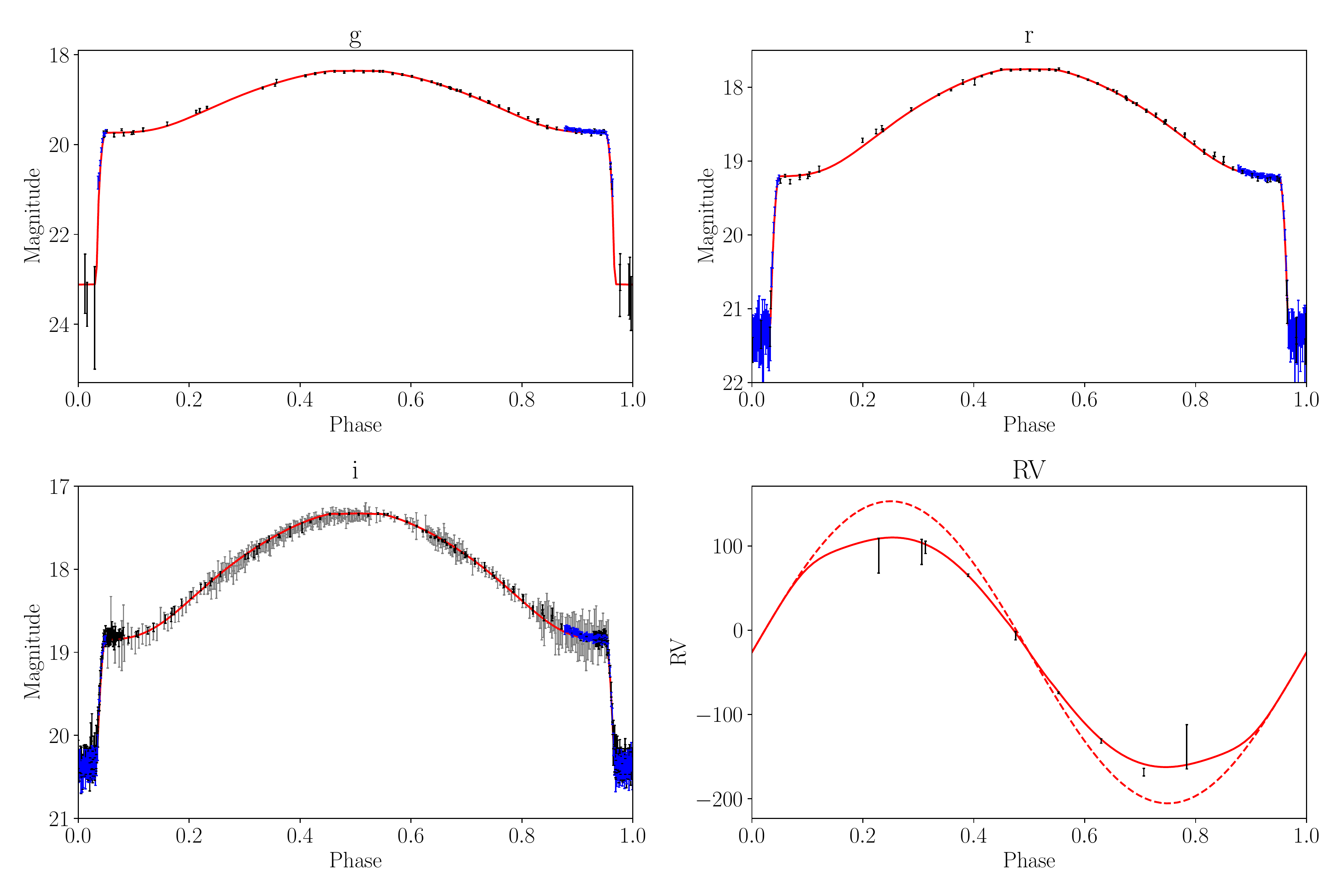}
    \caption{Phase folded light and RV curves of the central binary of Ou~5 overlaid on our best-fitting model curves computed with \textsc{phoebe 2.3}.  Points from the IAC80 are shown underlaid in grey, beneath the greater precision points from the INT-WFC and WHT-ACAM in black, while the binned WHT-HiPERCAM points are shown in blue.  The dynamical, centre-of-mass RV curve is shown as a dashed red line, while the centre-of-light RV curve is the solid red line.}
    \label{fig:model}
\end{figure*}

Time-series multi-band photometry was taken with $g$-, $r$- and $i$-band filters with the 2.5m Isaac Newton Telescope (INT) Wide Field Camera (WFC) on the nights 21-23 August 2015 and 1-5 August 2016 with integration times 90s, 120s and 90s, respectively.  Further multi-band photometry of the primary eclipse was obtained on the night of 17 October 2017 during first light of HiPERCAM on the WHT\footnote{ The instrument was still being optimised during this period which adversely affected the signal-to-noise of these data.} \citep{hipercam_wht,hipercam}, where the instrument was commissioned before its move to the 10.4m Gran Telescopio Canarias \citep[GTC;][]{dhillon18}.  Simultaneous exposures of 2.145s were taken in all bands\footnote{HiPERCAM is a five-band imager with "Super SDSS" filters of which we only use the $g_s$, $r_s$ and $i_s$, that are comparable to the standard Sloan filters in which we have data from other instruments that provide full phase coverage.} for a duration of approximately 1.5 hours beginning prior to ingress and continuing through egress (with approximately 8ms dead time between exposures).

All WFC and ACAM data were debiased and flat-fielded using standard routines of the astropy-affiliated python package \textsc{ccdproc} \citep{ccdproc}, while the other data were reduced using the respective instrument pipelines.  Differential aperture photometry using a constant aperture of radius 1.25 arcsec was then performed \citep[using the \textsc{photutils} package;][]{photutils} against the field stars IGAPSJ211422.28+434053.0 and IGAPSJ211423.16+434139.2 for the $i$-band, and against IGAPSJ211417.70+434148.0 and IGAPSJ211418.81+434109.5 for the $g$- and $r$-bands. This allowed the observations to be placed on an approximate apparent magnitude scale using the calibrated magnitudes of the comparison stars in the IGAPS catalogue \citep{igaps}. 
Due to the close proximity ($\lesssim$ 3 arcsec away) of a number of bright field stars to the South of the central star, nebular background subtraction was performed using a dedicated aperture, of the same size as the central star and comparison star apertures, shifted 3.5 arcsec Northeast of the central star.  Similarly, just as in \citet{Ou52014Corradi}, exposures taken under poor seeing conditions (in our case $>2$ arcsec) were discarded to avoid contamination in the central star aperture from nearby stars (as were HiPERCAM images where the central star was simply too faint to be measured -- hence the lack of g-band HiPERCAM photometry around mid-primary-eclipse).

The refined orbital ephemeris, determined using the $i$-band light curve (which has the largest number of data points as well as the longest temporal coverage), is 
\begin{equation}
\mathrm{HJD}_\mathrm{min}=2456597.49975(7) +  0.3642268(1) E 
\end{equation}

for the Heliocentric Julian Date of the mid-point of the primary eclipse (HJD$_\mathrm{min}$).  Both the reference time of mid-eclipse (T$_0$) and orbital period ($P_\mathrm{orb}$) are in reasonable agreement with those of \citet{Ou52014Corradi}, lying within approximately one uncertainty of their values.  The observed light curves from all instruments are presented folded on the refined ephemeris in Figure \ref{fig:model}.  The individual HiPERCAM exposures were of relatively poor signal-to-noise due to the extremely short exposure times, and hence the extracted photometry is shown binned into 500 equally spaced phase bins (and the same binning was applied before the modelling in Sec. \ref{sec:phoebe}).  

\begin{figure}
    \centering
    \includegraphics[width=\columnwidth]{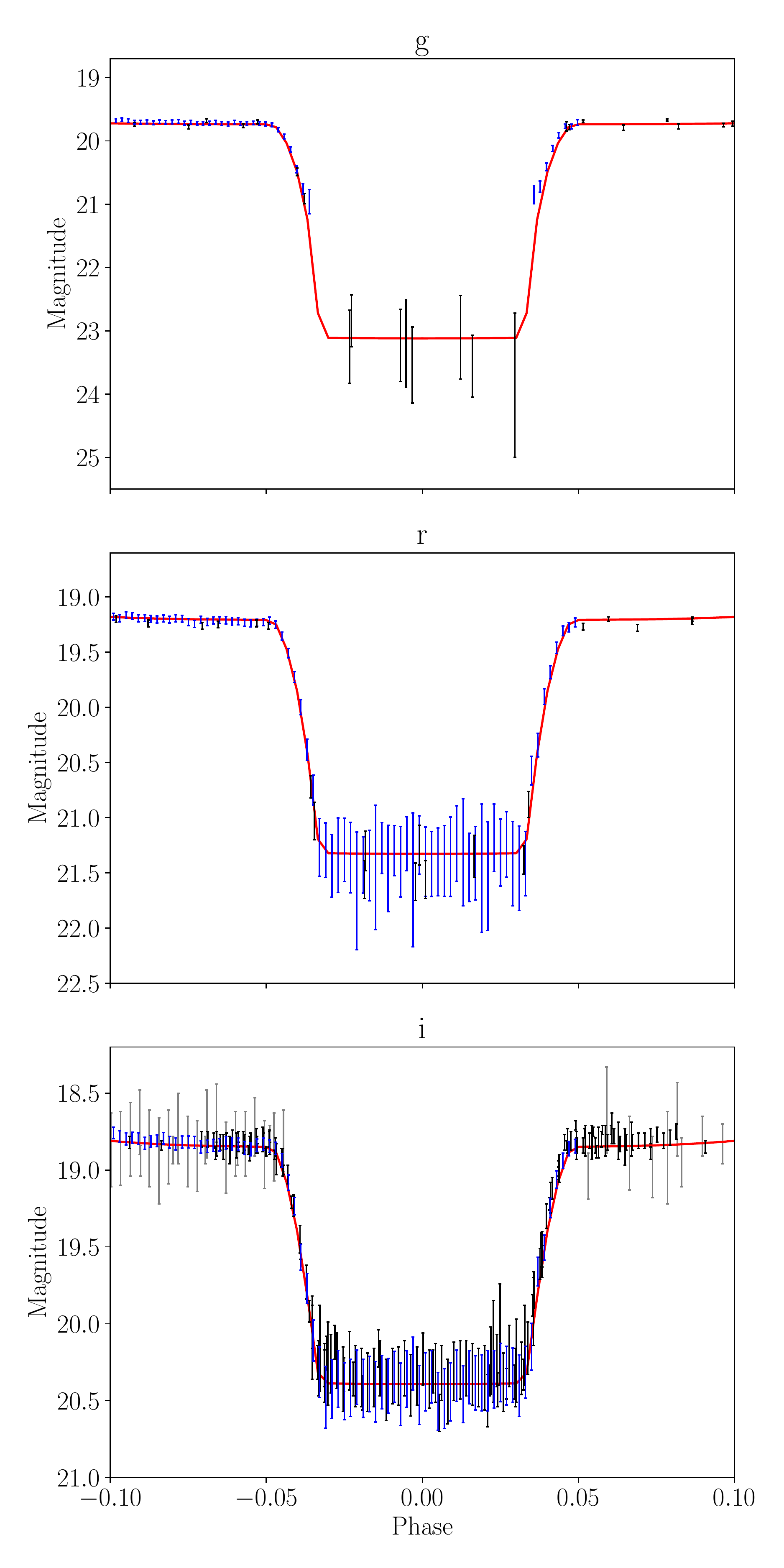}
    \caption{Phase folded light curves around primary eclipse of the central binary of Ou~5 overlaid on the best-fitting \textsc{phoebe 2.3} model (solid red line).  The colours of the points are as in Fig.\ \ref{fig:model}.}
    \label{fig:primeclipse}
\end{figure}

\begin{figure}
    \centering
    \includegraphics[width=\columnwidth]{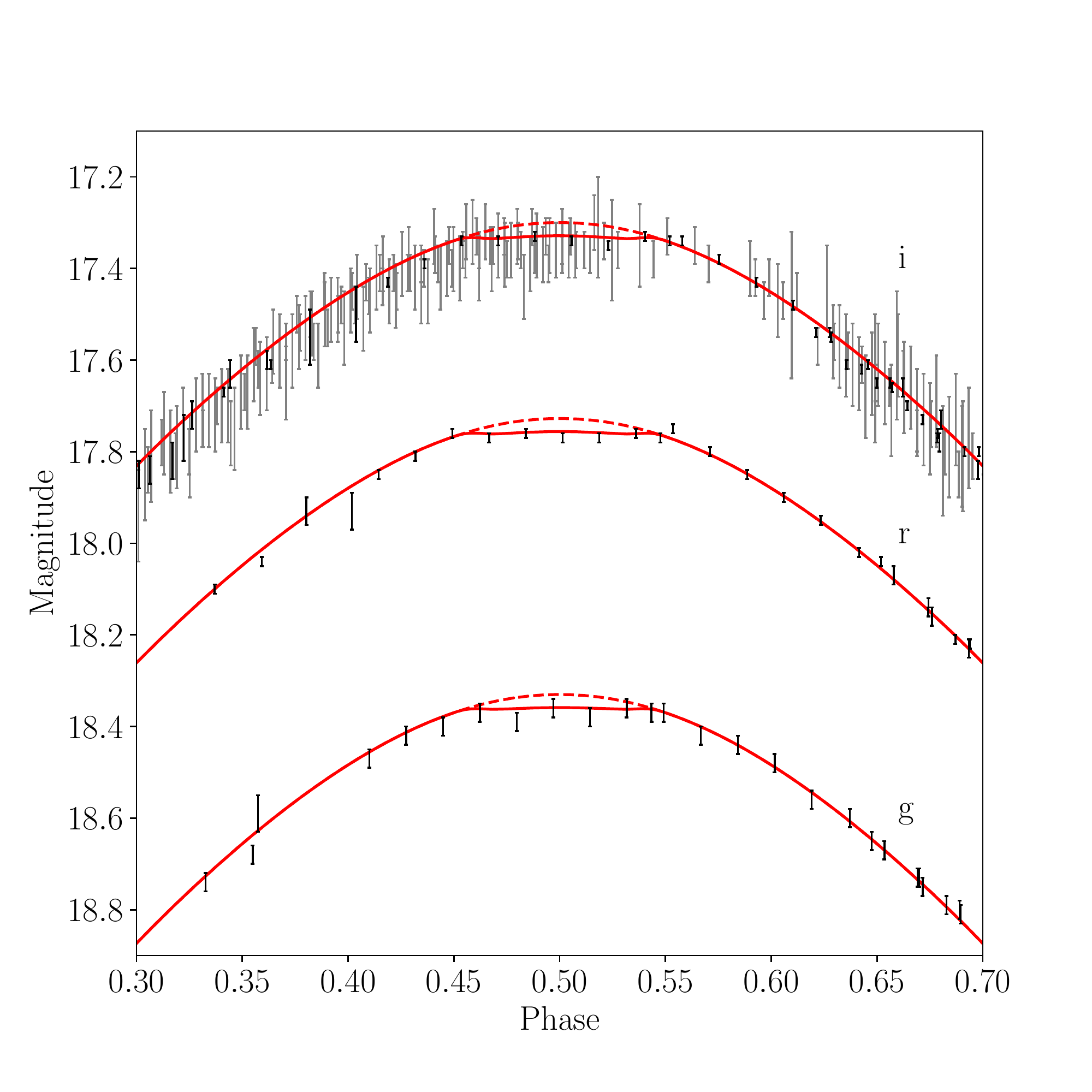}
    \caption{Phase folded light curves around secondary eclipse of the central binary of Ou~5 overlaid on the best-fitting \textsc{phoebe 2.3} model (solid red line) and the same model ignoring the secondary eclipse (dashed red line).  The colours of the points are as in Fig.\ \ref{fig:model}.}
    \label{fig:2ndeclipse}
\end{figure}

The light curves in each band present with a roughly similar morphology -- a large scale sinusoidal variability, which can be attributed to irradiation of a cool companion by the hot pre-CE primary, upon which deep primary (at phase $\phi=0$) and much shallower secondary ($\phi=0.5$) eclipses are superimposed.  The amplitude of irradiation effect is strongly passband-dependent, with a semi-amplitude of 0.675 mag, 0.725 mag and 0.75 mag in the $g$-, $r$- and $i$-bands, respectively.  The depth of the primary eclipse is also passband-dependent but, unlike the irradiation effect, is shallower in the redder bands with approximate depths of 3.6 mag, 2.2 mag and 1.5 mag in the $g$-, $r$- and $i$-bands, respectively (see Fig.\ \ref{fig:primeclipse}).  Evidence of a secondary eclipse is visible in all passbands with no strong indication for a passband-dependence, although the precision of the photometry is insufficient to accurately measure its depth (in a model independent way; see Fig.\ \ref{fig:2ndeclipse}) in any of the bands.  Both eclipses last approximately 47 min (0.090 in phase) with the primary eclipse being clearly flat-bottomed for more than 35 min (0.068 in phase) -- strongly indicative of a total eclipse.

\subsection{Spectroscopy}
\subsubsection{Radial velocities}
\label{sec:rvs}

The central binary of Ou~5 was observed with the blue-arm of the Intermediate-dispersion Spectrograph and Imaging System (ISIS) mounted on the WHT.  The R1200B grating was employed along with a 1 arcsec wide longslit resulting in a spectral resolution, $R\sim4700$, over the range 4000--4700\AA{}.  Seven exposures -- each of 2400s integration time -- were obtained on 17 August 2014 with a further two taken on 17 October 2014. The data were debiased, flat-fielded and wavelength calibrated with standard starlink routines \citep{figaro}.

The spectra were then continuum subtracted before cross-correlation against a model template comprising a flat continuum with the complex of irradiated emission lines (N~\textsc{iii} ${\lambda\lambda}$4634.14, 4640.64, C~\textsc{iii} ${\lambda\lambda\lambda}$4647.42, 4650.25, 4651.47\,\AA{} and C~\textsc{iv} ${\lambda}$4658.30) superimposed \citep[as in][]{jones20b,munday20}.  Unfortunately, as is frequently the case for relatively bright PNe \citep[e.g.,][]{miszalski11}, it was not possible to adequately subtract the bright and somewhat irregular nebular emission lines (namely those of the Balmer and Pickering series) in order to derive accurate radial velocities (RVs) for the hot component of the binary.  Similary, without being able to isolate the stellar absorption lines from the primary (due to the nebular contamination), we are unable to spectroscopically constrain the temperature and surface gravity of the nebular progenitor.

The RV measurements of the irradiated emission line complex, following heliocentric correction, are shown in Table \ref{tab:RVs}, while the data are shown folded on the ephemeris determined from the photometry in Figure \ref{fig:model}.  The RVs present a sinusoidal variability with phasing roughly consistent with the photometric ephemeris determined in Sec.\ \ref{sec:phot} (i.e., with a maximum at $\phi\sim$0.25 and minimum at $\phi\sim0.75$), although the phase coverage is not complete and the uncertainties increase dramatically away from the maximum of the irradiation effect at $\phi=0.5$, as the irradiated lines become weaker and the signal-to-noise decreases.  The measured RVs indicate a semi-amplitude, $K_\mathrm{irrad}$ for the region of the secondary from which the irradiated lines emanate, of approximately 130 \kms{}.  Noting that this is likely a minimum value for the centre-of-mass (CoM) semi-amplitude (as the irradiated line complex would be expected to be displaced towards the irradiating body), this implies a mass function $f(M_1)=\frac{{M_1}^3\sin^3i}{(M_1+M_2)^2}\gtrsim0.083$~M$_\odot$, where $M_1$ is the mass of the hot primary.

\begin{table}
\caption{Heliocentric radial velocity measurements of the irradiated emission line complex from the companion to the central star of Ou~5.}
\centering
\begin{tabular}{ r r l}
\hline
HJD & \multicolumn{2}{c}{RV (km~s$^{-1}$)} \\
\hline
2456887.538251 &     98.4    &    $\pm$  7.3\\
2456887.566322 &     65.2    &    $\pm$  1.5\\
2456887.597478 &     $-$6.5  &    $\pm$  5.0\\
2456887.625582 &    $-$74.3  &    $\pm$  1.1\\
2456887.653629 &   $-$131.5  &    $\pm$  2.6\\
2456887.681699 &   $-$168.5  &    $\pm$  4.5\\
2456887.709797 &   $-$138.4  &    $\pm$ 26.3\\
2456945.419667 &     88.5    &    $\pm$ 20.8\\
2456945.447906 &     92.9    &    $\pm$ 14.8\\
\hline 
\end{tabular}
\label{tab:RVs}
\end{table}

\subsubsection{Eclipse spectroscopy}
\label{sec:gtcspec}

The flat-bottomed nature of the primary eclipse is strongly indicative of a total eclipse, where all the light from the central binary observed at this phase should originate from the companion. As such, additional spectroscopy targeting these phases was obtained with the GTC and Optical System for Imaging and low-Intermediate-Resolution Integrated Spectroscopy (OSIRIS) using the R1000R grism and a 0.6 arcsec wide longslit (resulting in a resolution of $R\sim1000$ over the range 5\,000\,\AA{} $\lesssim\lambda\lesssim$ 10\,000\,\AA{}).
In total, six 25 min exposures (timed to begin and end during the 35 min flat-bottomed region of the primary eclipse) were obtained on the nights 8 August, 30 September \& 6 December 2016 and 3, 5 \& 10 August 2018. Data were reduced following standard reduction routines from \textsc{figaro} \citep{figaro} in the \textsc{starlink} software package \citep{starlink}, and flux-calibrated using observations of standard stars taken with the same instrumental set-up. The individual spectra were then combined and dereddened assuming an R$_\mathrm{V}$ of 3.1 and A$_\mathrm{V}$ of 2.0 \citep[as derived from the nebula in][]{Ou52014Corradi} following the wavelength-dependent extinction function of \cite{fitzpatrick99extinction}. 

\begin{figure}
    \centering
    \includegraphics[width=\columnwidth]{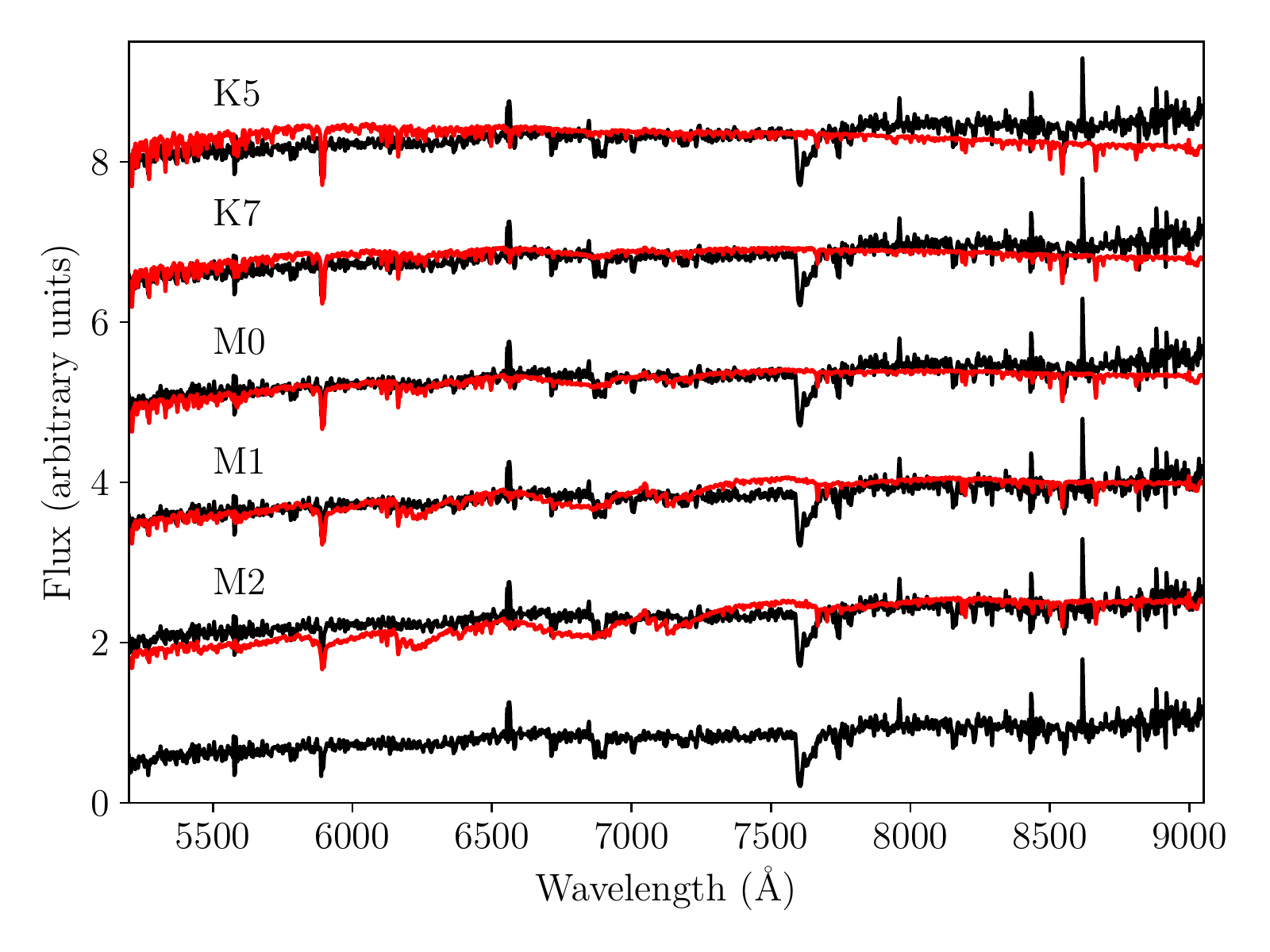}
    \caption{The combined spectrum during the primary eclipse (black), and then repeated and vertically shifted by an arbitrary constant with the most similar template spectra from \citet{PyHammer2020} overlaid (red). All template spectra have solar metallicity.  Note that the observed line emission is most likely under-subtracted nebular emission (e.g., H$\alpha$+[N~\textsc{ii}] around 6560~\AA{}), while the absorption features at $\sim$6900~\AA{} and  $\sim$7600~\AA{} are O$_2$ telluric bands.}
    \label{fig:M0M1M2}
\end{figure}

To our knowledge, this is the first time that an isolated spectrum of the companion to a post-CE PN central star has been obtained\footnote{\citet{miszalski13} present spectroscopy of the central star of the Necklace around photometric minimum, where the red-end of the spectrum is dominated by the companion.  However, as the central star of the Necklace is not eclipsing, some contribution from the hot pre-white-dwarf primary was inevitable.  Similarly, \citet{liebert95} obtained spectra of BE Uma (the central star of LTNF~1) during the ingress of primary eclipse, but with no flat bottom to the eclipse some contribution from the companion is again inevitable.}.  Unfortunately, the signal-to-noise ratio of the combined, dereddened spectrum is not sufficient to estimate the stellar parameters (surface gravity, effective temperature) via either spectral synthesis or equivalent widths.  However, the spectrum was compared to catalogue spectra with the \textsc{PyHammer} python package \citep{keseli17,roulston20}. This indicates that the companion is a late K-dwarf or an early M-dwarf (see Fig.\ \ref{fig:M0M1M2}), indicating an effective temperature of $\sim$4\,000~K \citep{cifuentes20}.  The metallicity of the comparison spectra was allowed to vary freely, along with the spectral type, but was not well constrained due to the relatively poor signal-to-noise of the combined Ou~5 central star spectrum.

The eclipse colours (see Fig.\ \ref{fig:primeclipse}) are similarly a useful probe of the spectral type of the companion.  The dereddened colour indices of the central star during eclipse, $g-r\sim$1.1 and $r-i\sim$0.6, are consistent with a late K-type or early M-type companion \citep{cifuentes20}, in good support of the findings of the GTC-OSIRIS spectroscopy.  The absolute magnitude of such a star should be roughly $M_i\sim$7.4 \citep{cifuentes20}, which corresponds to a distance, $D\sim$2.2 kpc, for the dereddened mid-eclipse apparent magnitude, $i$=19.1.  The H$\alpha$ surface-brightness-radius relationship of \citet{frew16}, however, gives a distance of 5.3$\pm$1.0 kpc, significantly farther away than the above estimate \citep[PNe with close binary central stars are known to lie away from the standar relation, but not to such a degree][]{frew08}.  Unfortunately, Gaia EDR3 contains a negative parallax ($\varpi=-0.013$ mas) for the central star of Ou~5 \citep{chornay21b}, although the uncertainty ($\sigma_\varpi=0.109$) is sufficiently large that the parallax distance is within roughly two standard deviations of the distance modulus.  \citet{bailerjones18} probablistically estimate a distance, from the Gaia parallax, of 3.6~kpc, with a confidence interval of 2.4--5.5~kpc.

\section{PHOEBE modelling}
\label{sec:phoebe}

In order to further probe the parameters of the central stars of Ou~5, we simultaneously modelled the light- and RV- curves (described in Sec.\ \ref{sec:phot} and \ref{sec:rvs}, respectively) using the \textsc{phoebe2} code \citep[][]{phoebe2,phoebe4,phoebe5}.  The light curves were modelled using the 'absolute' mode for passband luminosities \citep[i.e., the integrated passband fluxes were returned in absolute units without any internal rescaling; see Sec.\ 3.4.4 of][]{phoebe5} and were then placed on a Vega magnitude scale (to match the IGAPS calibration; see Sec.\ \ref{sec:phot}) using \textsc{phoebe2}-calculated fluxes of a model Vega \citep[with mass, temperature, radius, and distance set to match the values found by][]{vega}.  Thereby, the modelling accounts not only for the amplitude of variability in each band but also the difference in brightness between the bands (i.e., the colour as a function of phase).
Fitting was then performed via a Markov chain Monte Carlo (MCMC) method \citep[using \textsc{emcee};][]{emcee} parallelised on the LaPalma3 supercomputer \citep[as in, e.g.,][]{jones19,jones20b,munday20}.  

The mass ($M_1$), temperature ($T_1$), and radius ($R_1$) of the primary were allowed to vary freely, while its emergent spectrum was modelled using the T\"ubingen non-LTE model atmosphere package \citep[TMAP;][]{rauch03,werner03} model atmospheres \citep[for further details of the grid used and implementation in \textsc{phoebe2}, see][Jones et al.\ in prep]{reindl16}.  The secondary, on the other hand, was modelled using Castelli \& Kurucz model atmospheres \citep{ck2004}, and its mass ($M_2$), temperature ($T_2$) and radius ($R_2$) were restricted to values in very broad agreement with the spectral type derived in Sec. \ref{sec:gtcspec} (i.e., ranging from K0 through to M8).  \textsc{phoebe2}'s native `interpolated' limb-darkening was used for both stars \citep{phoebe2}.  As the light reprocessed by the secondary tends to be focussed into bright irradiated lines (rather than simply heating and re-radiation), the albedo of the secondary was allowed to vary between passbands, thus accounting for the presence of different numbers and strengths of irradiated lines in different bands.  Similarly, the albedo in each band was allowed to reach values in excess of the theoretically accepted bolometric range of 0.6--1.0, again to reflect the fact that the reprocessed light might be emitted significantly more strongly in some bands than others (i.e., while the net bolometric albedo must be unity or less, in some passbands it may be significantly greater).  The binary inclination ($i$) and systemic velocity ($\gamma$) were allowed to vary freely.

As discussed at length in \citet{jones20b} and \citet{munday20}, the RVs of the secondary (Sec.\ \ref{sec:rvs}) do not represent its CoM RV but rather that of the zone from which the irradiated lines emanate (which should thus be displaced towards the source of irradiation and exhibit a lower amplitude).  As such, the ``flux-weighted'' mode of RV measurement in \textsc{phoebe2} was employed, under the assumption that the irradiated centre-of-light (CoL) of the secondary would be a better approximation for the origin of the irradiated lines.

\begin{figure*}
\centering
\includegraphics[width=\textwidth]{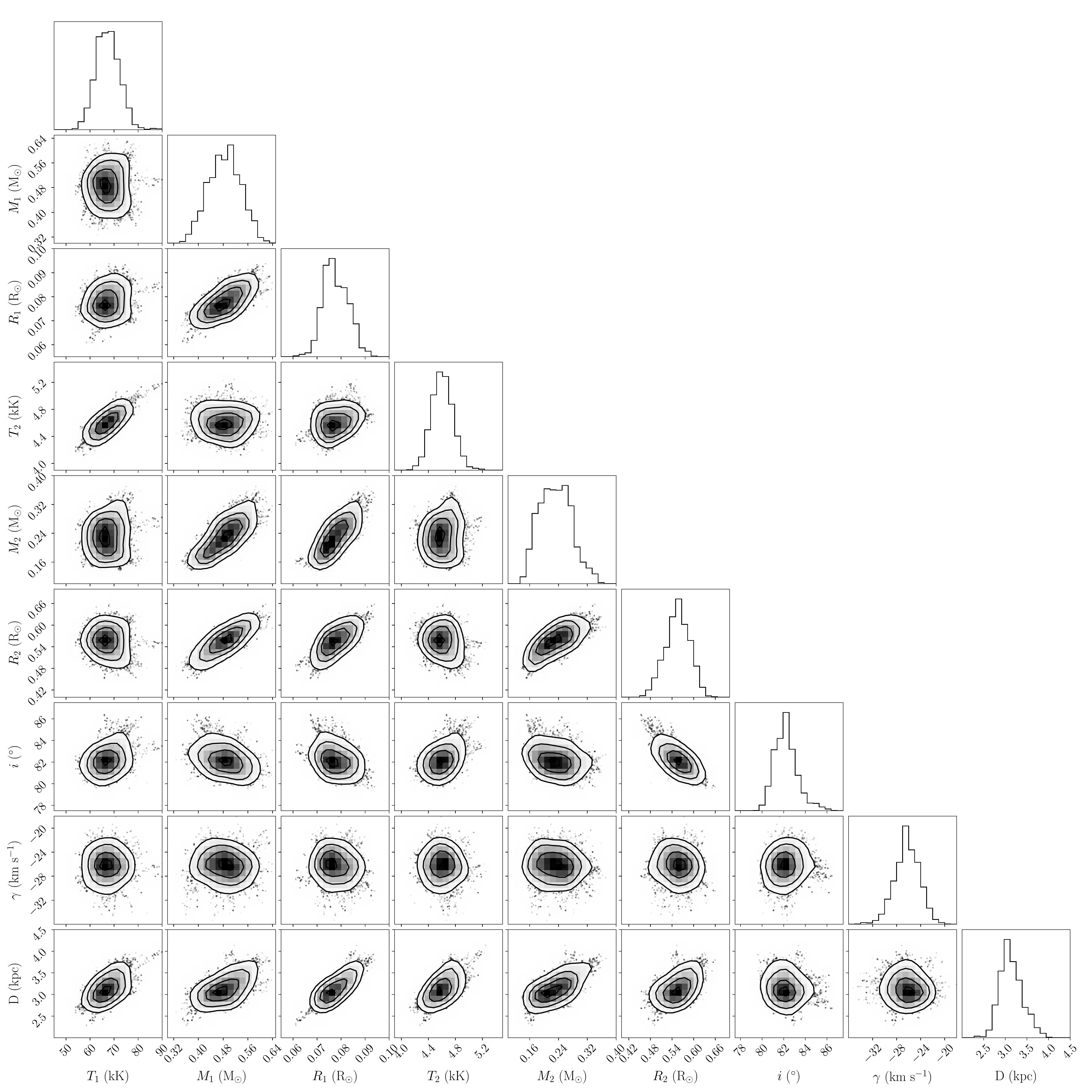}
\caption{A corner plot \citep[made using \textsc{corner};][]{corner} of the MCMC posteriors of the \textsc{phoebe}2-model parameters (see Tab.\ \ref{tab:CSparams} and the text for details).}
\label{fig:corner}
\end{figure*}

The final model light- and RV-curves are shown along with the data in Figure \ref{fig:model}, and the best-fitting parameters (the median of the MCMC posteriors) are listed in Table \ref{tab:CSparams} along with their 1$\sigma$ uncertainties (the 16$^{th}$ and 84$^{th}$ percentiles of the MCMC posteriors). A corner plot highlighting the parameter posterior distributions and their interdependencies is shown in Fig.\ \ref{fig:corner}.  The quality of the fit is, in general, very good, with residuals on the order of one standard deviation.  The synthetic light curves reproduce well both the depth of the primary eclipse and the overall shape of the irradiation effect, including around quadrature where the preliminary model of \citet{Ou52014Corradi} was poorest.  The synthetic RV curve displays a slight Rossiter-McLaughlin-like deviation from sinusoidality just as in \citet{jones20b} as a result of the transition of the photocentre from the irradiated face to the non-irradiated face -- coinciding with the phases at which no irradiated lines would be expected to be observed and thus at which the secondary RV would not be able to be measured.

\begin{table*}
\centering
\caption{Parameters of the central stars of Ou~5 as determined by the \textsc{phoebe}2 modelling using CoL RVs and interpolated limb-darkening for the secondary (see text for details).  Note that the uncertainties here are purely statistical and based on the MCMC posterior distributions (and, as such, are almost certaintly underestimated).}              
\label{tab:CSparams}      
\centering                                      
\centering
\begin{tabular}{llrlrl}
\hline\hline
&& \multicolumn{2}{c}{Hot component} & \multicolumn{2}{c}{Cool component} \\
\hline
\multicolumn{2}{l}{Mass (M$_\odot$)} & 0.50&$\pm$0.06 & 0.23&$^{+0.05}_{-0.03}$\\
\multicolumn{2}{l}{Radius (R$_\odot$)} & 0.078&$\pm$0.006 & 0.56&$^{+0.09}_{-0.07}$\\
\multicolumn{2}{l}{T$_\mathrm{eff}$ (kK)} & 67.2&$^{+4.9}_{-4.6}$ & 4.6&$\pm$0.2\\
& g-band & \multicolumn{2}{c}{1.0 (fixed)} & 0.67 &$\pm$0.11\\
Albedo & r-band & \multicolumn{2}{c}{1.0 (fixed)} & 0.78 &$\pm$0.11\\
& i-band & \multicolumn{2}{c}{1.0 (fixed)} & 0.97 &$\pm$0.14\\
\hline
\multicolumn{2}{l}{Orbital period (days)} & \multicolumn{2}{r}{0.3642268}&\multicolumn{2}{l}{$\pm$0.0000001}\\
\multicolumn{2}{l}{Orbital inclination} &\multicolumn{2}{r}{82.1$^\circ$}&\multicolumn{2}{l}{$^{+1.1^\circ}_{-1.0^\circ}$}\\
\multicolumn{2}{l}{Heliocentric systemic velocity (\kms{})} & \multicolumn{2}{r}{-26.2} &\multicolumn{2}{l}{$\pm2$}\\
\multicolumn{2}{l}{Distance (kpc)} & \multicolumn{2}{r}{3.1} &\multicolumn{2}{l}{$\pm0.3$}\\
\hline
\end{tabular}
\end{table*}

Before entering into a detailed discussion of the derived model parameters it is important to highlight the limitations that must be taken into account in their interpretation.  The quoted values and their uncertainties are based solely on the distribution of the MCMC posteriors (see Fig.\ \ref{fig:corner}).  These do not take into account any underlying uncertainties or systematic errors based on the modelling assumptions \citep[for example, the treatment of irradiation;][]{horvat19,jones20b}.  Similarly, the tightness of the posterior distribution may not always be indicative of a good fit, instead simply being the result of convergence to the best-fitting model. Therefore, the posteriors may not reflect the presence of any systematic deviations that can be symptomatic of flaws in the underlying modelling assumptions and/or the merit function prioritising certain light or RV curve features that more strongly constrain some parameters than others. Ultimately, the overall quality of the fit and the convergence of the posteriors serve as a strong indication of the validity of the modelling, but the aforementioned caveats must be considered before perhaps over interpreting the results or accepting the posterior distributions as being representative of the true uncertainties.

The temperature of the best-fitting model companion would imply a spectral type somewhat earlier than derived using the mid-eclipse spectroscopy in Sec.\ \ref{sec:gtcspec} (though still roughly consistent given the low signal-to-noise ratio of the spectra).  The derived secondary radius is well in line with that predicted by the mid-eclipse spectroscopy.  As no primary RVs could be extracted (see Sec.\ \ref{sec:rvs}), the secondary mass is only indirectly constrained and thus the posterior distribution is rather broad (see Fig.\ \ref{fig:corner}).  However, the derived mass (even accounting for uncertainties) is indicative of a later spectral type. Interestingly, models with more massive, and thus less Roche lobe filling (the best-fitting companion is $\sim$80 per cent Roche lobe filling), secondaries tended to more poorly reproduce the overall shape of the irradiation effect \citep[showing the same underestimation around quadrature as in the preliminary models of ][]{Ou52014Corradi}.

The model primary mass is on the cusp between post-RGB and post-AGB evolution \citep{hall13,mmmb}. The low model luminosity would certainly seem to imply a post-RGB evolution, but the model parameters are not entirely consistent with evolutionary tracks (see Fig.\ \ref{fig:kiel}) -- although this is not unusual,  with many post-CE central stars lying away from these tracks \citep{jones19,jones20b}.

Realistically, although the posterior distribution for the primary mass is relatively tight, the uncertainty should be appreciably larger simply because we do not know precisely from which region of the secondary the irradiated lines originate.  The modelling technique outlined above assumes that the locus of the irradiated lines is coincident with the irradiated CoL of the secondary.  Repeating the modelling assuming the CoM RV leads to a reduced mass for the primary of only 0.33$\pm$0.04~M$\odot$ -- definitively in a post-RGB regime.  The mass of the companion is also reduced in the CoM model but within the uncertainties of the CoL model, likewise the radii of both components are reduced, while the temperatures remain essentially the same.  Similarly, the model distance ($\sim$3~kpc) of both models is consistent with the distance estimates listed in Sec.\ \ref{sec:gtcspec}, particularly when one accounts for the possible post-RGB nature of Ou~5 (which might offer an explanation as to why the surface-brightness-radius relation of \citealt{frew16} would over-estimate the distance).

Ultimately, without RVs from the primary or a clearer idea of how the RVs of the irradiated lines compare to its CoM RV, the masses of both components remain rather poorly constrained.  However, the relatively invariant nature of the temperatures and radii between the models computed for CoM and CoL secondary RVs, highlights that these parameters are well constrained by the observations.  This is perhaps unsurprising as they are principally constrained by the depths and widths of the eclipse -- i.e., while the amplitude of the irradiation effect might well be reproduced with a different combination of primary temperature and secondary albedo, a larger primary temperature/luminosity would not be consistent with the observed depths of the primary eclipse.  In any case, the derived primary temperature ($T_1\sim$70~kK) lies on post-RGB evolutionary tracks for a mass of $\sim$ 0.4~M$_\odot$ (Fig.~\ref{fig:kiel}), perhaps indicative that the origin of the irradiated lines lies somewhere between CoL and CoM of the companion.

\section{Conclusions}
\label{sec:conc}

The properties of the post-CE binary central star of Ou~5 were probed via a combination of simultaneous modelling of multiband light- and RV-curves, and comparison of low-resolution spectra obtained during the primary eclipse with catalogue template spectra.

The secondary was found to be of late K- or early M-type from the eclipse spectroscopy but, while the light- and RV-curve modelling derives a radius consistent with this spectral type, the mass and temperature are more indicative of later and earlier spectral types, respectively.  This overall pattern of the secondary being both hotter and larger than expected given its mass is seemingly a typical trait of post-CE central stars, with the majority of systems demonstrating similar parameters \citep{demarco08,jones15,jones20b}.  Indeed, the derived companion mass and radius are similar to those found for Abell~65 \citep{hillwig15}, DS~1 \citep{hilditch96}, and, in particular, Abell~63 \citep{afsar08} -- although the modelled companion temperature is lower than derived in those systems (4.6~kK c.f.\ 5--6~kK).  

The heating and/or inflation observed in the companions of post-CE central stars is generally ascribed to the star not yet having thermally adjusted to a period of brief but intense accretion immediately prior to the CE \citep[e.g.,][and references therein]{jones15}.  Although, the extreme levels of irradiation experienced by the companion may also play a role \citep{demarco08}. Interestingly,  the companions of older post-CE binaries, where the primary has already reached the white dwarf cooling track, do not show evidence of inflation \citep[e.g.,][]{parsons18}, even though the Kelvin-Helmholtz timescale for these stars is generally too long for them to have relaxed back to a ``normal'' radius following the CE.  The models of \citet{prialnik85} show that for intermediate values of accretion efficiency and accretion rate (their figure 2), fully convective stars turn partially convective with the core contracting in response to the accretion while a second convective layer develops in the outer envelope.  It is this outer envelope which expands in response to the accretion, such that the radius and luminosity are roughly proportional to the total mass accreted.  This may, perhaps, offer a reason for why the more evolved post-CE companions are not inflated as, with only the outer envelope of the star expanding, it could feasibly have relaxed on an appreciably shorter timescale \citep[as is observed in the donors of cataclysmic variables;][]{stehle96}.

The primary star of Ou~5 is found to be low mass ($M_1\sim0.3-0.5$~M$_\odot$) and relatively cool ($T_1\sim70$ kK), inconsistent with single-star post-AGB evolutionary tracks.  The model primary's parameters are seemingly more in line with a post-RGB evolution -- an interesting prospect given that the nebular abundances (in particular the very low N/O) might be more consistent with nucleosynthetic yields following only the first dredge-up \citep[although the high He abundance would imply a second dredge-up phase and a massive progenitor;][]{karakas14} and that high ADFs have been speculatively associated with a post-RGB evolution \citep{jones16}. 

\begin{figure}
    \centering
    \includegraphics[width=\columnwidth]{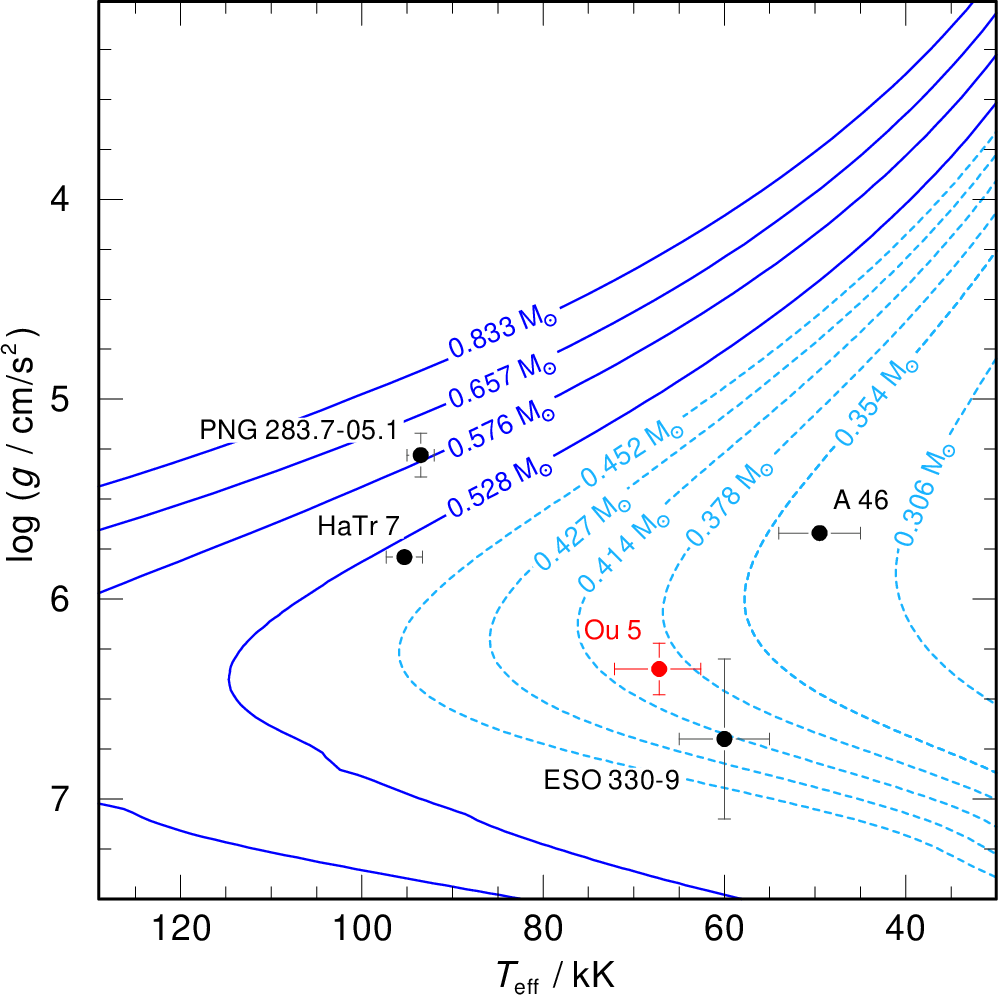}
    \caption{A Kiel diagram with the parameters of the strongest candidate post-RGB post-CE central stars known to date \citep[][and this work]{hillwig17,jones19}.  Underlaid are the post-AGB tracks of \citet[solid, dark blue]{mmmb} and post-RGB tracks of \citet[dashed, light blue]{hall13}.}
    \label{fig:kiel}
\end{figure}

Very few binary central stars of PNe have been demonstrated to be post-RGB rather than post-AGB.  \citet{hillwig17} identified five potentially post-RGB PNe to which \citet{jones20b} added one more, making the discovery of another candidate in Ou~5 of particular interest (the candidate systems for which effective temperatures and surface gravities have been determined are shown on a Kiel diagram in Fig.~\ref{fig:kiel}).  Previously, there were doubts as to whether post-RGB systems would produce observable PNe, but both these observational findings and the theoretical work of \citet{hall13} have clearly demonstrated that post-RGB PNe do indeed exist.  However, where dynamical masses have been measured, they are often at odds with the remnant mass which would be predicted based on post-RGB evolutionary tracks -- for example,  PN~G283.7$-$05.1 has a mass (as described from combined light and radial velocity curve modelling) which is lower than would be estimated from evolutionary tracks \citep[indeed, its effective temperature and surface gravity are far more consistent with post-AGB evolutionary tracks][]{jones20b}, while Abell~46 \citep{pollacco94,afsar08} and Ou~5 (this work) are ``too massive''. 

As a significant fraction of (naked) post-CE white-dwarf-main-sequence binaries are found to be post-RGB systems \citep[roughly one third;][]{rebassa-mansergas11}, one would therefore expect to find that many post-CE central stars are also post-RGB systems. The characterisation and discovery of further such systems will ultimately be key in constraining the importance and properties of this pathway for PNe, in general.

\section*{Acknowledgements}

The authors would like to thank the referee, Orsola De Marco, for her insightful review of the manuscript.  

DJ acknowledges support from the Erasmus+ programme of the European Union under grant number 2020-1-CZ01-KA203-078200. JG-R acknowledges support from the Severo Ochoa excellence program  CEX2019--000920--S. JG-R and RLMC acknowledge support from the Canarian Agency for Research, Innovation and Information Society (ACI-ISI), of the Canary Islands Government, and the European Regional Development Fund (ERDF), under grant with reference ProID2021010074. DJ, JG-R and RLMC acknowledge support under grant P/308614 financed, by funds transferred from the Spanish Ministry of Science, Innovation and Universities, charged to the General State Budgets and with funds transferred from the General Budgets of the Autonomous Community of the Canary Islands by the MCIU. JM acknowledges the support of the ERASMUS+ programme in the form of a traineeship grant, and STFC in the form of a studentship. PS acknowledges financial support by the Polish NCN grant 2015/18/A/ST9/00578.  VSD and HiPERCAM were funded by the European Research Council under the European Union’s Seventh Framework Programme (FP/2007-2013) under ERC-2013-ADG Grant Agreement no.\ 340040 (HiPERCAM) and the STFC.

This paper is based on observations obtained with:  the 2.5-m~Isaac Newton (INT) and 4.2-m~William Herschel (WHT) telescopes of the Isaac Newton Group of Telescopes; the 10.4~m Gran Telescopio Canarias (GTC) and the 0.8~m IAC80 telescope operating on the islands of La Palma and Tenerife at the Spanish Observatories of the Roque de Los Muchachos and Teide of the Instituto de Astrof\'\i sica de Canarias.

This research made use of computing time available on the high-performance computing systems at the Instituto de Astrof\'isica de Canarias. The authors thankfully acknowledge the technical expertise and assistance provided by the Spanish Supercomputing Network (Red Espa\~nola de Supercomputaci\'on), as well as the computer resources used: the LaPalma Supercomputer, located at the Instituto de Astrof\'isica de Canarias.  The authors also acknowledge support from the Agencia Estatal de Investigaci\'on del Ministerio de Ciencia e Innovaci\'on (AEI-MCINN) under grant reference 10.13039/501100011033.

\section*{Data Availability}

All raw IAC80-CAMELOT, INT-WFC, WHT-ACAM, WHT-ISIS and GTC-OSIRIS data are available from the respective online archives, while the WHT-HiPERCAM data is available upon reasonable request to the authors.  All extracted photometry and radial velocities are available in the article or from VizieR at the CDS.



\bibliographystyle{mnras}
\bibliography{Ou5} 



\bsp	
\label{lastpage}
\end{document}